\documentclass[sigconf]{acmart}

\usepackage{booktabs}
\usepackage{hyperref}
\usepackage{enumitem}
\usepackage{xspace}
\usepackage{subcaption}
\usepackage{needspace}

\usepackage{soul} 
\newcommand{\canadaHidden}{UofT\xspace}
\newcommand{\nlHidden}{TU Delft\xspace}

\AtBeginDocument{  }

\copyrightyear{2026}
\acmYear{2026}
\setcopyright{cc}
\setcctype{by}
\acmConference[ITiCSE 2026]{Proceedings of the 31st ACM Conference on Innovation and Technology in Computer Science Education V. 1}{July 10--15, 2026}{Madrid, Spain}
\acmBooktitle{Proceedings of the 31st ACM Conference on Innovation and Technology in Computer Science Education V. 1 (ITiCSE 2026), July 10--15, 2026, Madrid, Spain}
\acmDOI{10.1145/3803400.3809346}
\acmISBN{979-8-4007-2634-7/2026/07}

\begin{document}

\title[AI-Generated Traces for Novice Programmers]{AI-Generated Traces for Novice Programmers: Learning Effects and Learner Differences in a Multi-Institutional Study}



\author{Yuri Noviello}
\orcid{0009-0006-5846-7756}
\affiliation{  \institution{Delft University of Technology}
  \city{Delft}
  \country{Netherlands}
}
\email{y.noviello@tudelft.nl}
\author{Naaz Sibia} \orcid{0000-0001-7628-7077}
\affiliation{
\institution{University of Toronto}
   \city{Toronto}
   \country{Canada}
}
\email{naaz.sibia@utoronto.ca}

\author{Anastasiia Birillo}
\orcid{0000-0003-2269-8211}
\affiliation{  \institution{JetBrains Research}
  \city{Belgrade}
  \country{Serbia}
}
\email{anastasia.birillo@jetbrains.com}

\author{Thomas Overklift Vaupel Klein} \orcid{0009-0002-3606-3620}
\affiliation{  \institution{Delft University of Technology}
  \city{Delft}
  \country{Netherlands}
}
\email{t.a.r.overkliftvaupelklein@tudelft.nl}

\author{Michael Liut} \orcid{0000-0003-2965-5302}
\affiliation{
\institution{University of Toronto Mississauga}
   \city{Mississauga}
   \country{Canada}
}
\email{michael.liut@utoronto.ca}

\author{Gosia Migut}
\orcid{0000-0002-4120-5454}
\affiliation{  \institution{Delft University of Technology}
  \city{Delft}
  \country{Netherlands}
}
\email{m.a.migut@tudelft.nl}

\renewcommand{\shortauthors}{Yuri Noviello et al.}


\begin{abstract}
Introductory programming (CS1) courses often struggle to support students’ understanding of program execution. While visualizations can make execution processes explicit, their effectiveness depends on design and context, and empirical evidence for AI-generated visualizations remains limited.
We propose Generated Animated Traces (\textit{GATs}), AI-generated, analogy-based, narrated animations that coordinate source code, execution state, and conceptual analogies. We conduct a study at two institutions in CS1 courses (Python, $N=961$; Java $N=151$) comparing \textit{GATs} to textual explanations. We measure immediate learning performance and experience, end-of-course engagement and exam performance.
Results show that \textit{GATs} can yield selective benefits for immediate learning, but benefits are context-dependent and short-term. We observe that \textit{GATs}' influence on performance is moderated by learner engagement profiles. 
This finding underscores the importance of personalized approaches.
\end{abstract}

\maketitle
\vspace{6em}

\begin{figure*}[h]
    \centering
    \includegraphics[width=0.8\linewidth]{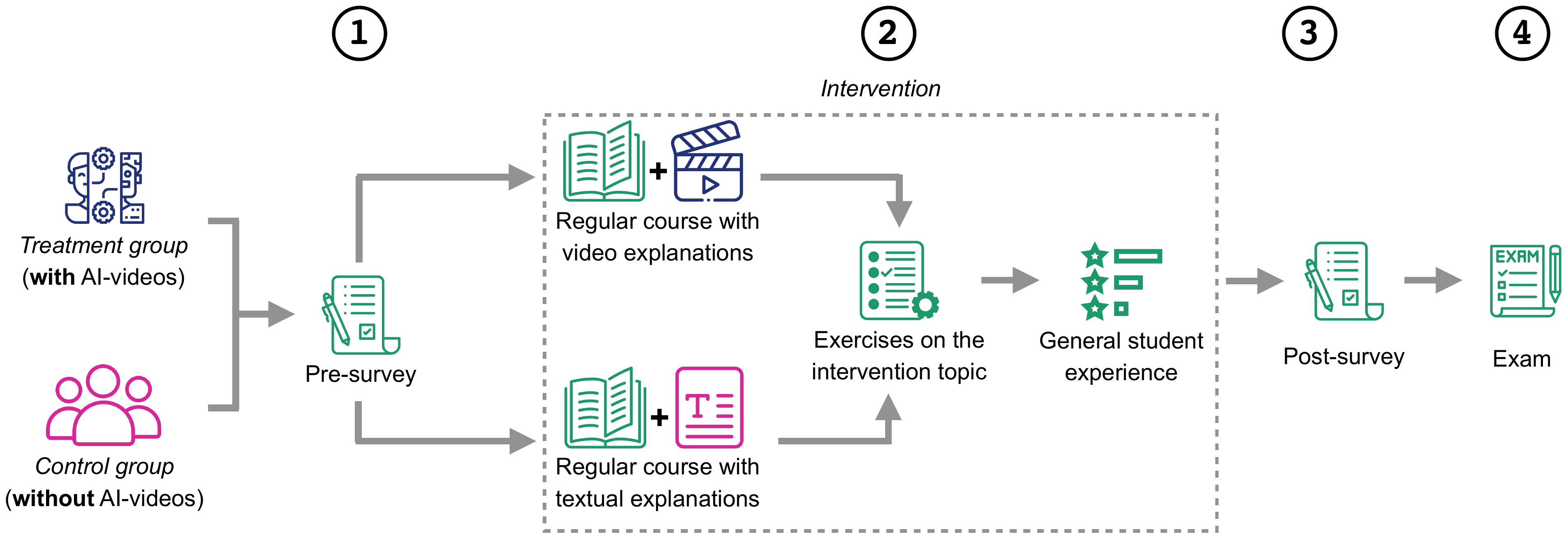}
     \vspace{-0.3cm}
    \caption{Study pipeline. Unique materials for the \textit{treatment group} are marked in blue, and unique materials for the \textit{control group} are marked in pink. Shared materials for both groups are marked in green.     }
    \label{fig:methods:pipeline}
\end{figure*}

\section{Introduction}
\label{sec:introduction}

Understanding how programs execute is a central challenge in introductory programming (CS1) courses. Novice programmers may learn to produce correct outputs while still lacking robust mental models of how variables, memory, and control flow interact at runtime, which limits transfer and long-term learning ~\cite{sibia2025codeconceptevaluatingmultiple}.

A common instructional response is to make program execution more explicit through visual explanations, such as traces, diagrams~\cite{cunningham2017using,thomas2004scaffolding}, animations that present multiple views of execution~\cite{sorva2013review}, or domain-related analogies~\cite{saxena2023achieving}. Prior work suggests that coordinating verbal and non-verbal channels can support comprehension~\cite{dual-coding}, while dynamic visualizations with attention guidance help learners focus on task-relevant elements~\cite{de2017attention}. Advances in Large Language Models (LLMs) now make it possible to automatically generate such instructional materials at scale, creating new opportunities to deploy rich visual explanations consistently across large-enrollment courses~\cite{anvil, review-AI-video}. 
However, the educational impact of such materials remains unclear. Empirical evidence is limited on when, for whom, and under what conditions AI-generated instructional materials support learning, particularly across institutional contexts and learner profiles~\cite{review-AI-video}. Prior work further suggests that instructional interventions rarely benefit all learners equally, and that average effects may obscure meaningful differences in engagement and learning outcomes~\cite{guo2018non}.

In this work, we investigate Generated Animated Traces (\textit{GATs}), an AI-generated, analogy-based, narrated animation that synchronizes a code snippet, a memory diagram, and a conceptual analogy within a single presentation to explicitly represent how program state and control flow evolve during execution. Rather than assuming benefits from coordinated views, we empirically evaluate how GATs compare to carefully matched step-by-step textual explanations across two CS1 courses with different languages and course structures.
We address three research questions:

\begin{enumerate}[leftmargin=1.22cm,start=1,label={\bfseries RQ\arabic*:}]
        \item How do \emph{GATs} affect (a) immediate post-intervention programming task performance and (b) course summative performance, compared to textual explanations? 
            \item How does students’ learning experience differ between \emph{GATs} and textual explanations, as measured by self-reported frustration, cognitive load, and situational interest?
    \item To what extent are the effects of \emph{GATs} moderated by learner engagement profiles?
\end{enumerate}

In this work we provide the first systematic evaluation of GATs for CS1 across: (1) two institutions with different languages and structures, (2) multiple instruments (immediate/summative performance, learning experience, engagement), and (3) learner heterogeneity via engagement profiles. Our findings show that the impact of \textit{GATs} is context- and learner-dependent, underscoring the need to move beyond average effects in CS1 instruction.

\vspace{-1.8em}
\section{Related Work}
\label{sec:related}

Learning to program requires more than producing correct outputs. It requires building mental models that link code to runtime behavior, including variable state, control flow, and memory operations~\cite{johnson1983mental, sorva2013review}. In CS1, instruction is commonly scaffolded through notional machines, idealized models of how programming languages execute, that help learners reason about scope, state changes, and execution order~\cite{fincher2020notional}. Effective learning, therefore, depends on supporting abstraction rather than surface-level code tracing alone.

Program visualizations are widely used instructional tools for making program execution explicit. Tools such as Python Tutor expose step-by-step execution traces and runtime state~\cite{guo2013online}, which can support tracing accuracy. However, highly detailed visualizations may anchor novices to low-level mechanics, limiting abstraction and transfer~\cite{sorva2013review, eckert2022loops}. Learners may successfully follow traces without developing deeper conceptual understanding.

However, learners do not automatically integrate multiple representations when exposed to program visualizations. In an eye-tracking study, novices often attend selectively to a single view unless explicitly guided~\cite{bednarik2012expertise}. Synthesizing this literature, Sorva concludes that visualization effectiveness depends not on visual richness alone, but on how representations are coordinated~\cite{sorva2013review}. 
Similarly, to avoid redundancy and cognitive overload, the DeFT framework emphasizes that multiple external representations require careful design~\cite{ainsworth2006deft} and should guide attention to task-relevant elements~\cite{sweller2011cognitive, mayer2008increased}.
Also, the expertise-reversal effects suggest that instructional supports that are helpful for novices may hinder more advanced learners~\cite{kalyuga2007expertise}. These findings underscore that adding instructional detail does not inherently improve learning outcomes.

In addition to program visualizations, analogies are frequently used in CS1 to ground abstract execution concepts in familiar domains. Well-designed analogies have been shown to reduce cognitive load and improve conceptual understanding~\cite{forivsek2012metaphors, cao2016examining}. However, their effectiveness is highly design-sensitive, and misaligned analogies can mislead learners or impede abstraction~\cite{bettin2022semaphore}. At the same time, the analogy-based videos may increase engagement even when performance gains are limited~\cite{zhu2025comparing}, highlighting a distinction between learning experience and measured outcomes. 

As an extension of visualization- and analogy-based instruction, recent studies indicate that AI-generated teaching materials, when pedagogically sound, can be as effective as human-produced materials~\cite{netland2025comparing, xu2025recorded}, although they tend to score lower on social presence~\cite{review-AI-video}. Also, AI can assist learners in generating personalized analogies~\cite{bernstein2024nesting, harper2024tool}. As AI lowers the cost of producing rich instructional media for various needs, there is growing interest in deploying such materials at scale~\cite{review-AI-video}. However, empirical evidence remains limited on when, for whom, and under what conditions AI-generated instructional materials support learning, particularly across institutional contexts and learner profiles. Addressing this gap requires evaluations that move beyond average effects to consider learner heterogeneity and multiple learning outcomes.

\vspace{-1em}
\section{Methods}
\label{sec:methods}

\subsection{Study design}

\textbf{Overview.} 
In 2025, we conducted two similar studies at \textit{Delft University of Technology} (\nlHidden) and the \textit{University of Toronto Mississauga}
(\canadaHidden), both research-intensive publicly funded institutions. Students at both sites were enrolled in an undergraduate introductory programming course (CS1), using Java at \nlHidden and Python at \canadaHidden, and were randomly assigned to either \textit{treatment} (AI-generated videos) or \textit{control} (textual explanations) groups (Figure~\ref{fig:methods:pipeline}). 
~\\
\textbf{Instruments.} At the start of the course, students completed a pre-survey on \textit{baseline programming experience} (Figure~\ref{fig:methods:pipeline} \textbf{(1)}). Both groups participated in interventions on selected topics (Figure~\ref{fig:methods:pipeline} \textbf{(2)}). Immediately following each intervention students completed topic-aligned exercises, used to measure \textit{short-term learning performance} (Figure~\ref{fig:methods:pipeline} \textbf{(2)}). The exercises included programming task (contributing 80\% to the immediate performance) and 2 MCQs (multiple-choice questions, each contributing 10\% to the immediate performance). One MCQ was about code tracing (identify output of a given program), another one was on conceptual understanding (describe execution of a given program). Students also answered Likert-scale questions about \textit{self-reported cognitive load, frustration, and situational interest}, adapted from established self-report instruments (Paas mental effort, NASA-TLX, and IMI items)~\cite{paas1992,HartStaveland1988,IMI} (Figure~\ref{fig:methods:pipeline} \textbf{(2)}). 
At the end of the course students completed a post-survey assessing \textit{self-reported engagement and elaboration strategies}, using CAP (Passive, Active, Constructive) engagement instruments~\cite{ICAP,CAP} and MSLQ elaboration~\cite{msql} (Figure~\ref{fig:methods:pipeline} \textbf{(3)}). 
\textit{Long-term summative performance} was assessed via the final exam, using the same questions across institutions, adapted only for programming language (Figure~\ref{fig:methods:pipeline} \textbf{(4)}).
~\\
\textbf{Reproducibility}. All study materials (surveys, videos, explanations and exercises) are included in the replication package ~\footnote{Replication package: \url{https://osf.io/kb9uv}}.
~\\
\textbf{Data Privacy \& Ethics}. Ethical approval was obtained from both institutions. To ensure instructional equivalence, the control group received textual explanations identical in content to those used in the treatment videos. Students were informed about the study purpose and data usage practices. 
To protect privacy, all datasets were anonymized prior to analysis.

\subsubsection{Study 1 at \nlHidden}
The study took place in a 9-week CS1 course in the first quarter, taken by incoming CS majors and bridge-to-master’s students. Taught in Java, it covered core topics (e.g., control flow, methods, OOP, file I/O, streams, unit testing) via lectures and live coding. Students completed 8 optional weekly labs, practice exams (with up to 5\% bonus), and were assessed through a theoretical exam (50\%) and a programming exam (50\%). Interventions were embedded in labs, with voluntary participation incentivized by a raffle of ten €25 gift cards.

\subsubsection{Study 2 at \canadaHidden}
The study was conducted in a first-semester CS1 course for first-year students entering a competitive CS major, with an interdisciplinary cohort. The 12-week, Python-based course used a flipped classroom~\cite{bishop2013flipped} and covered core topics (e.g., control flow, functions, OOP, file I/O, sorting, complexity, regular expressions). Assessment included a term test (20\%), final exam (35\%), eleven weekly labs (3\% each), and LMS-based preparatory exercises (12\%). Interventions were mandatory within exercises, with a 2\% bonus for completing pre- and post-surveys.

\begin{figure}[t!]
    \centering
    \includegraphics[width=0.45\textwidth]{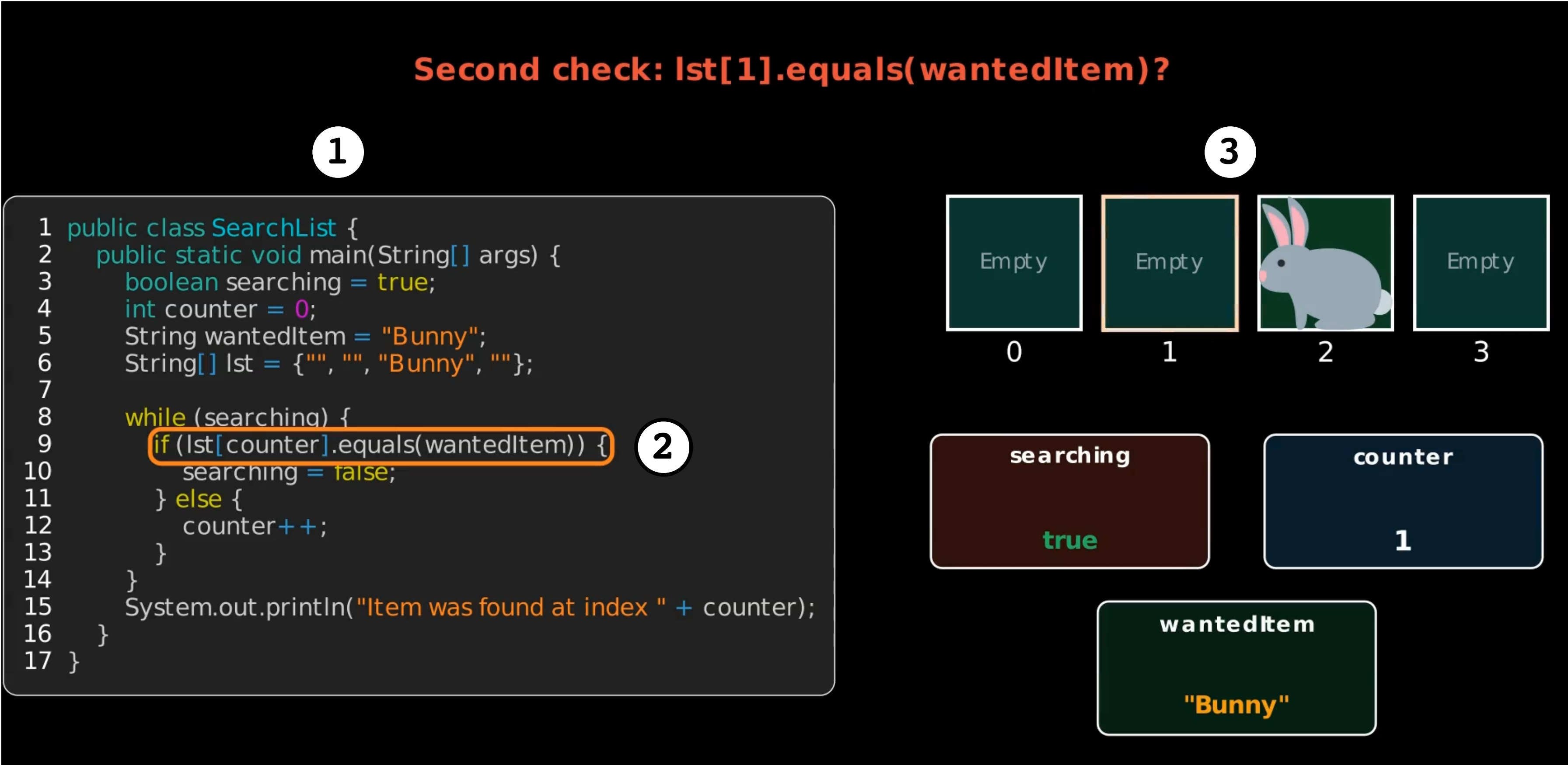}
     \vspace{-0.3cm}
    \caption{Example of a GAT provided to the treatment group: (1) a code snippet; (2) a highlighted code line indicating the program state to visualize; (3) a conceptual analogy to explain the highlighted code line.}
    \label{fig:methods:example}
\end{figure}
\begin{figure}[t!]
\centering
    \includegraphics[width=0.5\textwidth]{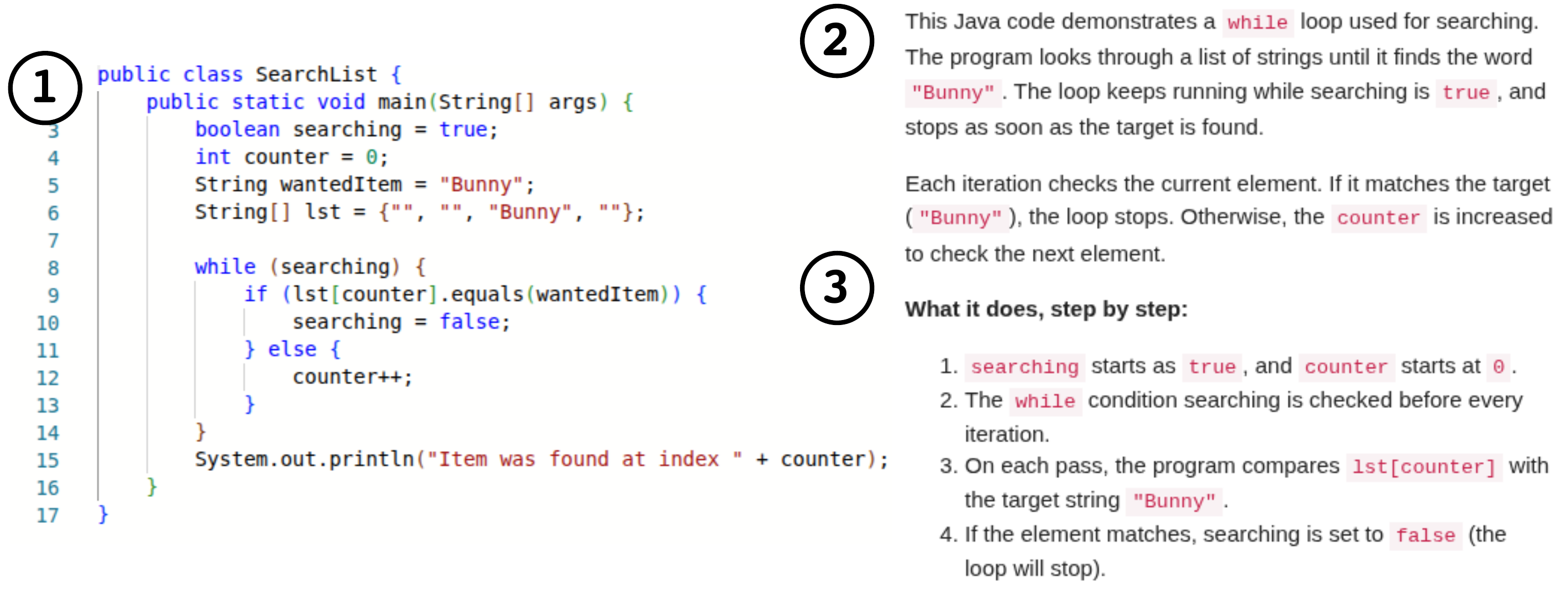}
             \caption{Example of a textual explanation provided to the control group: (1) the code snippet; (2) a general description of the code; (3) step-by-step execution order. The figure shows only the first few steps due to space limitations.}
    \label{fig:methods:weblab}
\end{figure}

\subsubsection{Intervention topics}
We selected five (5) topics for the interventions: \textit{while-loop}, \textit{arraylist}, \textit{hashmap}, \textit{file reading}, and \textit{file writing}.
These topics were selected because they are foundational CS1 concepts taught in both courses, aligning with CS2023 guidelines, but are also known to have common misconceptions among students~\cite{GarciaCraig2025}. For each topic, we had a matching test/exam question.
In study 1 at \nlHidden, we omitted \textit{file writing} in the final exam due to logistical issues.
In study 2 at \canadaHidden we deployed a selection of three (3) interventions: \textit{while-loop}, \textit{file reading}, and \textit{file writing}. Unfortunately, \textit{arraylist} and \textit{hashmap} had to be removed for two reasons.
First, the \canadaHidden deployment occurred in a larger, more heterogeneous population with mandatory participation, raising concerns about participant burden and engagement quality with extensive intervention materials. Second, practical constraints in course scheduling and assessment alignment made it challenging to meaningfully integrate all five topics within the existing curriculum structure. We selected topics that: (1) represented core CS1 concepts, (2) aligned with available exam items, and (3) were most likely to reveal effects given the population characteristics.

\subsection{Interventions}
For each topic an intervention consisted of an instructional artifact: video explanations (\textit{GATs}) for the treatment group and \textit{textual explanations} for the control group. The underlying code snippets and instructional logic were consistent across both groups, thus the intervention manipulated only the \textit{presentation modality}.

\subsubsection{Generated Animated Traces (GATs)}

A GAT is an AI-generated, analogy-based, narrated animation designed to support representational fluency by synchronizing two distinct views: 
~\\
\textbf{Source Code View} (Figure~\ref{fig:methods:example} \textbf{(1)}) with the currently executed line(s) highlighted (Figure~\ref{fig:methods:example} \textbf{(2)}) to guide student attention; 
~\\
\textbf{Conceptual Analogy View} that uses a visual analogy to foster abstraction (e.g., searching for a \textit{bunny} in physical boxes to illustrate a \textit{while} loop)  (Figure~\ref{fig:methods:example} \textbf{(3)}). The analogy view is augmented with variable value updates, data structure contents, and control-flow transitions directly within the analogical context.

GATs were produced through a human in the loop pipeline.
Initial scripts and execution traces were generated by an AI-driven system, that produces analogy-oriented explanations and automatically renders them as manim-based animations~\cite{anvil}. Subtitles for all movements were generated automatically, and a voiceover was added to provide audio narration. 
Final videos were lightly edited by the first author to refine timing, correct layout issues, and ensure pedagogical clarity and alignment with course instruction.
\subsubsection{Textual explanations}
Textual explanations consisted of a structured walkthrough of the \emph{same code} used in the corresponding GAT (Figure~\ref{fig:methods:weblab}).
Each explanation included the code snippet (Figure~\ref{fig:methods:weblab} \textbf{(1)}), a high-level description of its purpose (Figure~\ref{fig:methods:weblab} \textbf{(2)}), and a step-by-step execution narrative describing actions and resulting state changes (Figure~\ref{fig:methods:weblab} \textbf{(3)}).
To ensure comparability, textual explanations were matched exactly to GATs in learning objectives, conceptual scope, and line-by-line execution order.
\subsection{Data Analysis}
\label{sec:data_analysis_plan}
Our analysis plan was designed to estimate the effects of explanation modality on students' immediate and longer-term learning performance and experience.
We analyzed \nlHidden{} and \canadaHidden{} data separately. We did not pool data across institutions because the deployments differ in programming language and topic coverage.
~\\
\textbf{Design Integrity and Attrition Checks.}
We used independent t-tests to check baseline equivalence in prior programming experience and Mann–Whitney U tests to assess differential attrition and modality-related dropout bias~\cite{manntest}.
~\\
\textbf{Modeling Immediate Impact and Experience.}
Immediate performance outcomes were analyzed using \textit{Linear Mixed Models}~\cite{mixed-model} with student-level random intercepts to account for repeated interventions per student.
Models included \textit{Topic} and prior experience as covariates. We estimated pooled and topic-specific effects and applied this approach to immediate performance and experience instruments (cognitive load, frustration, and situational interest).
~\\
\textbf{Long-Term Performance and Psychometric Modeling.}
Exam outcomes were analyzed at the student level using \textit{Analysis of Covariance (ANCOVA)}~\cite{ANCOVA}, controlling for prior experience. Post-survey instruments (MSLQ elaboration and CAP engagement dimensions) were analyzed using analogous models to assess how GATs influenced students' end-of-course cognitive strategies.
~\\
\textbf{Engagement Profiles Clustering.}
To address learner heterogeneity, we conducted an exploratory engagement profile analysis using \textit{k-means} clustering on self-reported CAP instruments, in line with prior engagement profile analyses~\citet{engagment-profile}. This allowed us to test if the efficacy of GATs was moderated by a student's specific \textit{engagement profile}, moving beyond simple average effects.
~\\
\textbf{Error Control and Significance Reporting.}
Due to the large number of dependent variables, we controlled the \textit{False Discovery Rate (FDR)} using the \textit{Benjamini-Hochberg (BH)} procedure~\cite{benjamini1995controlling} within predefined instruments domains. We report effect estimates, standard errors, and both raw and adjusted $p$-values.

\section{Results}
\label{sec:results}
To validate the experimental design, we performed sanity checks on randomization and retention. For both institutions, no significant differences were found between the treatment and control groups for baseline equivalence (using self-reported prior programming experience) and differential attrition (via the number of completed interventions). 
The results of the analysis are presented in (Table~\ref{tab:big_table}).

\newcounter{rownumber}
\newcommand{\rownumber}{\stepcounter{rownumber}\therownumber}

\setlength{\tabcolsep}{1pt}

\begin{table*}[t]
\centering
\small
 \begin{tabular*}{\textwidth}{@{\extracolsep{\fill}}p{0.3cm}ll @{\hspace{4em}} rrrr @{\hspace{4em}} rrrr}
\toprule
 & Instrument & Topic / Context
& $N$ & $\Delta$ & $p$ & $p_{adj}$
& $N$ & $\Delta$ & $p$ & $p_{adj}$ \\
\midrule

\multicolumn{3}{l}{\textbf{Topic-specific effects}} & 
\multicolumn{4}{c}{\hspace{-10mm}\textbf{\nlHidden{}}} & 
\multicolumn{4}{c}{\textbf{\canadaHidden{}}} \\
\midrule
\rownumber & Immediate performance & \textit{While-loop}
& 138 & 8.227 & \textbf{0.003} & \textbf{0.015}
& 927 & -0.118 & 0.880 & 0.879 \\
\rownumber &Immediate performance & \textit{ArrayList}
& 99 & 1.347 & 0.651 & 0.651
& --- & --- & --- & --- \\
\rownumber &Immediate performance & \textit{HashMap}
& 52 & 5.771 & 0.090 & 0.218
& --- & --- & --- & --- \\
\rownumber &Immediate performance & \textit{File Reading}
& 34 & 4.270 & 0.296 & 0.370
& 764 & -0.762 & 0.378 & 0.566 \\
\rownumber &Immediate performance & \textit{File Writing}
& 20 & 7.933 & 0.131 & 0.218
& 416 & 2.691 & \textbf{0.021} & 0.062 \\
\addlinespace[0.4em]

\rownumber &Frustration & \textit{While-loop}
& 138 & 0.073 & 0.769 & 0.993
& 927 & -0.163 & \textbf{0.043} & 0.129 \\
\rownumber &Frustration & \textit{ArrayList}
& 99 & 0.006 & 0.982 & 0.993
& --- & --- & --- & --- \\
\rownumber &Frustration & \textit{HashMap}
& 52 & 0.003 & 0.992 & 0.993
& --- & --- & --- & --- \\
\rownumber &Frustration & \textit{File Reading}
& 34 & -0.542 & 0.139 & 0.484
& 764 & -0.086 & 0.331 & 0.331 \\
\rownumber & Frustration & \textit{File Writing}
& 20 & -0.610 & 0.194 & 0.484
& 416 & -0.158 & 0.164 & 0.246 \\
\addlinespace[0.4em]

\rownumber &Cognitive load & \textit{While-loop}
& 138 & 0.087 & 0.632 & 0.676
& 927 & -0.121 & 0.073 & 0.220 \\
\rownumber &Cognitive load & \textit{ArrayList}
& 99 & 0.150 & 0.439 & 0.676
& --- & --- & --- & --- \\
\rownumber &Cognitive load & \textit{HashMap}
& 52 & -0.090 & 0.676 & 0.676
& --- & --- & --- & --- \\
\rownumber &Cognitive load & \textit{File Reading}
& 34 & -0.299 & 0.241 & 0.657
& 764 & -0.040 & 0.582 & 0.582 \\
\rownumber &Cognitive load & \textit{File Writing}
& 20 & -0.354 & 0.263 & 0.657
& 416 & -0.132 & 0.159 & 0.239 \\
\addlinespace[0.4em]

\rownumber &Situational interest & \textit{While-loop}
& 138 & 0.218 & 0.257 & 0.367
& 927 & -0.038 & 0.619 & 0.849 \\
\rownumber &Situational interest & \textit{ArrayList}
& 99 & 0.219 & 0.291 & 0.367
& --- & --- & --- & --- \\
\rownumber &Situational interest & \textit{HashMap}
& 52 & 0.161 & 0.489 & 0.489
& --- & --- & --- & --- \\
\rownumber &Situational interest & \textit{File Reading}
& 34 & 0.472 & 0.091 & 0.367
& 764 & -0.016 & 0.849 & 0.849 \\
\rownumber &Situational interest & \textit{File Writing}
& 20 & 0.374 & 0.294 & 0.367
& 416 & -0.211 & \textbf{0.042} & 0.125 \\

\midrule
\multicolumn{10}{l}{\textbf{Overall immediate effects (pooled across topics)}} \\
\midrule
\rownumber  & Immediate performance & \textit{Pooled}
& 151 & 5.420 & \textbf{0.003} & \textbf{0.017}
& 961 & 0.173 & 0.760 & 0.761 \\
\rownumber & Frustration  & \textit{Pooled}
& 151 & -0.069 & 0.676 & 0.974
& 961 & -0.135 & \textbf{0.043} & 0.086 \\
\rownumber & Cognitive load  & \textit{Pooled}
& 151 & 0.004 & 0.974 & 0.974
& 961 & -0.095 & 0.097 & 0.129 \\
\rownumber  & Situational interest & \textit{Pooled}
& 151 & 0.240 & 0.079 & 0.196
& 961 & -0.059 & 0.362 & 0.362 \\

\midrule
\multicolumn{10}{l}{\textbf{Long-term performance }} \\
\midrule
\rownumber & Exam performance & \textit{While-loop}
& 135 & 0.746 & 0.390 & 0.780
& 893 & 0.808 & 0.369 & 0.945 \\
\rownumber & Exam performance & \textit{ArrayList}
& 98 & 1.656 & 0.201 & 0.780
& --- & --- & --- & --- \\
\rownumber & Exam performance & \textit{HashMap}
& 52 & 0.151 & 0.698 & 0.915
& --- & --- & --- & --- \\
\rownumber & Exam performance& \textit{File Reading}
& 33 & 0.011 & 0.915 & 0.915
& 763 & 0.005 & 0.945 & 0.945 \\
\rownumber & Exam performance & \textit{File Writing}
& --- & --- & --- & ---
& 416 & 0.011 & 0.917 & 0.945 \\

\midrule
\multicolumn{10}{l}{\textbf{Post-survey outcomes}} \\
\midrule
\rownumber  & Constructive engagement & \textit{Post-survey}
& 16 & 0.460 & 0.369 & 0.443
& 398 & 0.212 & \textbf{0.016} & \textbf{0.049} \\
\rownumber & Active engagement & \textit{Post-survey}
& 16 & 0.891 & 0.072 & 0.144
& 398 & 0.139 & 0.125 & 0.125 \\
\rownumber & Passive engagement & \textit{Post-survey}
& 16 & 0.743 & \textbf{0.019} & 0.076
& 398 & 0.136 & 0.058 & 0.087 \\
\rownumber & MSLQ elaboration & \textit{Post-survey}
& 16 & 0.202 & 0.292 & 0.438
& 398 & 0.054 & 0.488 & 0.488 \\
\bottomrule
\end{tabular*}
\caption{
Summary of intervention effects for \nlHidden{} and \canadaHidden{}.
$\Delta$ denotes the estimated difference between conditions (\textit{GAT} -- Text).
\textit{N} is the number of students included in each analysis after excluding invalid responses.
$p_{adj}$ values are FDR-adjusted within each site and outcome family.
Bold indicates $p < .05$ and $p_{adj}<.05$.
--- indicates not available for that site.}
\vspace{-2em}
\label{tab:big_table}
\end{table*}

\subsection{Immediate learning performance}
\textbf{Study 1 at \nlHidden:} \textit{GATs} significantly improved immediate performance across topics (Table~\ref{tab:big_table}, row 21), mainly driven by the \textit{While-loop} intervention (Table~\ref{tab:big_table}, row 1), where the treatment group outperformed the control group. Other topics showed positive but non-significant differences that failed correction.
~\\
\textbf{Study 2 at \canadaHidden:} No significant overall differences in immediate performance were found between conditions. Topic-specific results also showed no effects after correction, though \textit{File Writing} had the largest raw difference favoring \textit{GATs} (Table~\ref{tab:big_table}, row 5).

\subsection{Immediate learning experience}
\textbf{Study 1 at \nlHidden:} No reliable differences were found for frustration or cognitive load, across topics. Situational interest showed a weak positive trend favoring \textit{GATs}, but it did not survive correction. Topic-specific analyses found no significant effects, with the largest difference in situational interest on \textit{File Reading}.
~\\
\textbf{Study 2 at \canadaHidden:} \textit{GATs} reduced frustration in raw tests for pooled experience measures, but this did not survive FDR correction (Table~\ref{tab:big_table}, row 22). No reliable differences were found for cognitive load or situational interest (Table~\ref{tab:big_table}, rows 23–24). At the topic level, raw tests showed lower frustration for \textit{While-loop} (Table~\ref{tab:big_table}, row 6) and reduced situational interest for \textit{File Writing} (Table~\ref{tab:big_table}, row 20), but neither remained significant after correction.

\subsection{Long-term exam outcomes}
Across both \textbf{\nlHidden} and \textbf{\canadaHidden}, ANCOVA models controlling for prior experience showed no significant condition differences on summative exam outcomes, with all adjusted $p$-values exceeding conventional significance thresholds (Table~\ref{tab:big_table}, rows 25-29).

\subsection{Post-survey scales}

\textbf{Study 1 at \nlHidden{}:} The post-survey analysis yielded no significant effects (Table~\ref{tab:big_table}, rows 30-33). Raw analyses suggested higher passive engagement (Table~\ref{tab:big_table}, row 32) under \textit{GATs}, but this pattern was not robust after FDR correction.
~\\
\textbf{Study 2 at \canadaHidden:} \textit{GATs} produced a reliable increase in \textit{Constructive engagement} (Table~\ref{tab:big_table}, row 30). No other engagement instruments showed effects surviving correction, and \textit{MSLQ elaboration} did not differ between conditions (Table~\ref{tab:big_table}, row 31-33).

\begin{figure}[t!]
    \centering
    \includegraphics[width=0.49\textwidth]{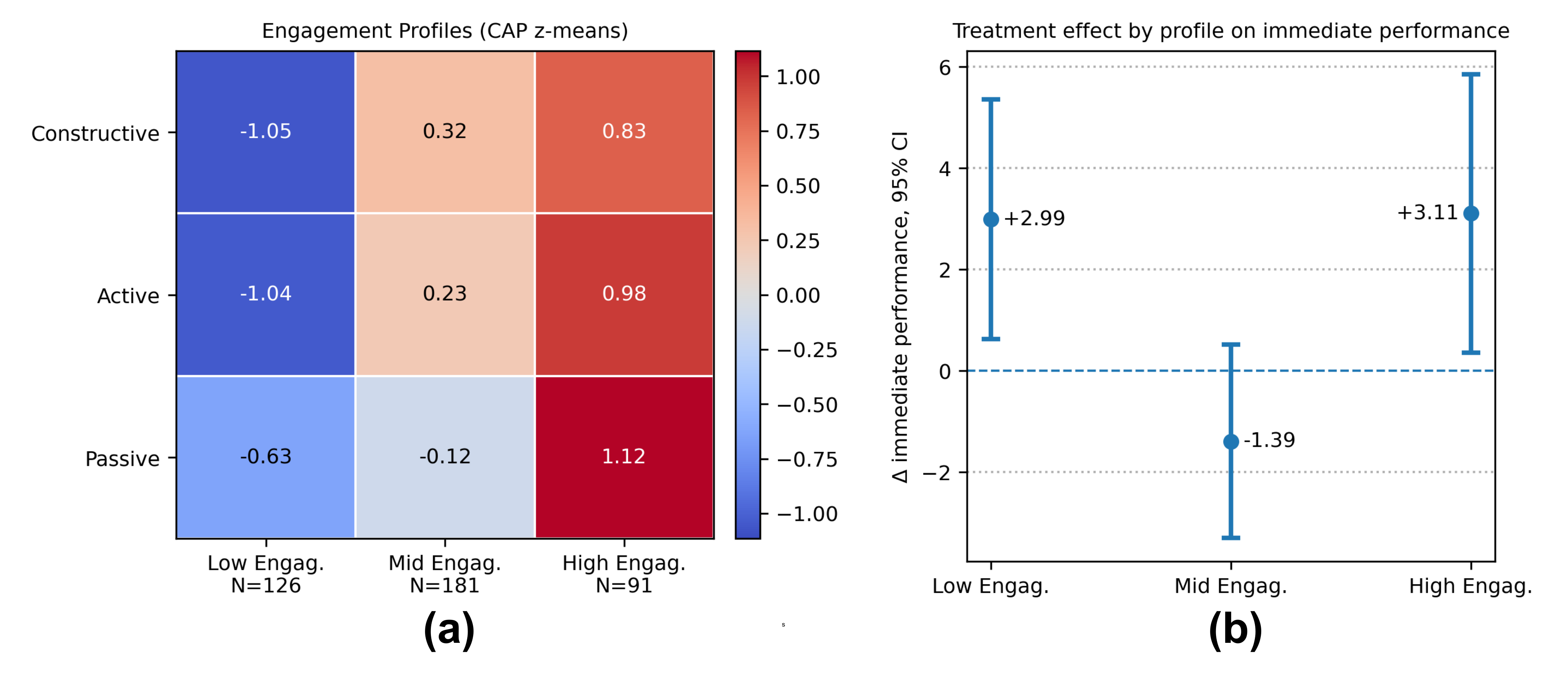}
    \vspace{-8mm}
    \caption{Engagement profiles and moderated treatment effects. (a) Heatmap shows mean standardized CAP engagement scores per cluster. (b) Within-profile treatment effects on immediate performance with 95\% confidence intervals.}
    \label{fig:persona-perfomance-plot}
\end{figure}

\subsection{Engagement profiles analysis}
We conducted an exploratory engagement profile analysis based on students' CAP engagement scores. Due to the limited sample size at \nlHidden{} ($N=16$), the analysis was conducted only for \canadaHidden{}.

We applied $k$-means clustering ($k=3$) to standardized CAP engagement dimensions (Passive, Active, Constructive), yielding three profiles that differed primarily in overall engagement magnitude, with each profile exhibiting consistently low-, mid-, or high-engagement scores~(Figure \ref{fig:persona-perfomance-plot} \textbf{(a)}). These engagement profiles were used to test whether the effects of \textit{GATs} varied across them.
We found no evidence of profiles' moderation for exam outcomes, cognitive load, frustration and situational interest.
In contrast, profiles significantly moderated immediate learning performances ($p=.005$, $p_{adj}=.024$). 
Specifically, \textit{GATs} benefited low- and high-engagement profiles, while mid-engagement profile showed a small performance decrement relative to textual explanations (Figure~\ref{fig:persona-perfomance-plot}~\textbf{(b)}). No other profile interactions were observed. 

\section{Discussion}
\label{sec:discussion}
\subsection{Effects on Learning Performance}
Our findings show that AI-generated visual explanations: (1) do not harm short and long-term performance and (2) can improve immediate performance, but that these benefits are context-dependent and short-term. 
In the Java-based course, \textit{GATs} produced a significant improvement in immediate performance when pooled across topics, driven primarily  
by the \textit{While-loop} intervention.
This suggests that GATs are particularly effective for conceptually demanding topics where learners must reason about invisible execution processes, such as iteration and state change. 
In contrast, no reliable pooled immediate performance effects were observed in the Python-based course. Although some topic-level differences favored \textit{GATs} in raw analyses, these effects did not survive correction.
This divergence highlights the importance of context-specific factors, including programming language, participation structure, and course organization. Prior work suggests that Java may impose higher syntactic and semantic overhead for novices than Python, potentially increasing intrinsic cognitive load and creating greater opportunity for representational scaffolds to provide benefit~\cite{Koulouri2014TOCE,Lokkila2023JISE, Kaila2023KoliCalling}.

Additionally, at \canadaHidden{}, participation was required (and grade-linked), which may dilute the effects if some students engage minimally with the materials.
At \nlHidden{}, participation was optional, which may yield a smaller but more engaged sample.

Across both institutions, we found no evidence that GATs influenced long-term performance. This pattern aligns with prior CS education work showing that isolated instructional interventions often improve local understanding without reliably transferring to delayed assessments~\cite{no-long-gain}. 
Repeated integration of visual explanations may be necessary to influence long-term outcomes.

\subsection{Effects on Experience and Engagement}
\textit{GATs} effects on immediate learning experience were modest compared to textual explanations. Across both deployments, pooled effects on frustration, cognitive load, and situational interest were small and did not survive correction. This may reflect the strength of the control condition: textual explanations were carefully constructed to provide a clear, step-by-step execution narrative, leaving limited room for additional reductions in perceived difficulty.

However, post-survey results paint a more nuanced picture.
In the larger study at \canadaHidden{}, \textit{GATs} produced a reliable increase in \textit{Constructive engagement} at the end of course, while Passive and Active engagement and MSLQ elaboration showed no effect.
Constructive engagement reflects sense-making behaviors such as generating inferences and connecting representations~\cite{CAP}. 
This suggests that the primary experiential benefit of \textit{GATs} may not lie in making tasks feel easier in the moment, but rather in shaping how students engage with explanations over time.
The absence of consistent immediate experience effects alongside the increase in Constructive engagement may suggest that while \textit{GATs} may not substantially reduce frustration or cognitive load beyond a strong textual baseline, they may still encourage deeper interpretive strategies, particularly in large, heterogeneous courses where sustained engagement is difficult to support at scale.

\subsection{Moderation by Engagement Profiles}
An important finding of this work is that average effects may mask meaningful learner differences.
Our exploratory analysis suggested that engagement profiles significantly moderated the effects of \textit{GATs} on immediate performance.

Low- and high-engagement students benefited from \textit{GATs} in immediate performance, while mid-engagement students showed a small performance decrement relative to textual explanations. 
Arguably, \textit{GATs} serve different functions depending on learners' baseline engagement profiles. For low-engagement students, \textit{GATs} 
may act as an \textit{attention-guided scaffold}, making key state transitions explicit and reducing reliance on self-generated tracing. 

In contrast, mid-engagement students who may already extract a coherent execution narrative from structured text, may experience redundancy or coordination costs when additional visual channels are introduced. This pattern is consistent with expertise-reversal effects observed in instructional design, where added scaffolding can hinder learners who no longer need it~\cite{kalyuga2007expertise, expertise-reversal-multimedia}.

Although exploratory and based on self-report measures, these findings suggest the importance of personalization. A single representation is unlikely to benefit all learners equally, and adaptive delivery based on, for example, engagement profiles may be crucial to realize the full potential of automated visual explanations.

\subsection{Limitations}
Our findings should be interpreted in light of several limitations.
First, the two deployments differed in programming language, participation incentives, and topic coverage; accordingly, we do not draw direct cross-institutional conclusions.
Second, the interventions targeted a limited set of topics, and observed effects did not transfer to long-term exam performance, suggesting more sustained integration.
Third, the engagement-profile moderation analysis was exploratory and relied on self-report engagement measures, so these results should be treated as suggestive rather than confirmatory.

\section{Conclusion}
\label{sec:conclusion}

This study evaluated the effect of AI-generated instructional visualizations to support novice programmers across two institutional contexts. Comparing \textit{GATs} with carefully matched textual explanations, we examined their impact on intermediate and long-term performance, learning experience, and engagement.
Our results show that \textit{GATs} can yield selective benefits for immediate learning, but these effects are context-dependent (topic, institution) and do not transfer to long-term exam performance. 

At the same time, in one institution, \textit{GATs} were associated with increased constructive engagement at the end of the course, and exploratory analysis revealed that learners' engagement profiles moderated the effectiveness of \textit{GATs} for the immediate learning performance.
Combining these findings suggests that the value of AI-generated visualizations lies not in uniformly improving outcomes, but in supporting specific learners and instructional moments. Future work should investigate adaptive, learner-aware strategies and examine whether personalized use of \textit{GATs} can lead to more durable effects on learning outcomes.    

\section{Acknowledgments}
This work was conducted as part of the AI for Software Engineering (AI4SE) collaboration between JetBrains and Delft University of Technology. The authors gratefully acknowledge the financial support provided by JetBrains, which made this research possible. We acknowledge the support of the Natural Sciences and Engineering Research Council of Canada (NSERC) Discovery Grant \#RGPIN-2024-04348 and PGS D–600673–2025.

\balance
\bibliographystyle{ACM-Reference-Format}
\bibliography{sample-base}

\end{document}